\documentclass[letterpaper,twocolumn,10pt]{article}
\usepackage{usenix,epsfig}

\usepackage{graphics}
\usepackage{color}
\usepackage[hyphens]{url}

\usepackage{breakurl}
\usepackage[breaklinks]{hyperref}
\usepackage{mathtools}

\usepackage[T1]{fontenc}

\usepackage[compact]{titlesec}

\usepackage{pdfpages}
\usepackage{amssymb,amsmath,amsthm}
\usepackage{enumitem}

\usepackage{varwidth}
\usepackage{placeins}

\usepackage{multirow}

\makeatletter
\def\imod#1{\allowbreak\mkern10mu({\operator@font mod}\,\,#1)}
\makeatother

\begin{document}
%\begin{CJK}{UTF8}{gbsn}
%\thispagestyle{fancy}
%\chead{\Large{This paper appeared in the Proceedings of the 19th USENIX Security Symposium, 2010}} 
%\pagestyle{empty}

\newcommand {\colline} [1] {\multicolumn{1}{|#1|}}
\DeclareUrlCommand\urltt{\urlstyle{tt}}

\date{}
\title{When Textbook RSA is Used to Protect the Privacy of Hundreds of Millions of Users}

%%for single author (just remove % characters)
\author{
{\rm Jeffrey Knockel}\\
Dept.\ of Computer Science\\
University of New Mexico\\
jeffk@cs.unm.edu
\and
{\rm Thomas Ristenpart}\\
Cornell Tech\\
ristenpart@cornell.edu
\and
{\rm Jedidiah R. Crandall}\\
Dept.\ of Computer Science\\
University of New Mexico\\
crandall@cs.unm.edu
% copy the following lines to add more authors
% \and
% {\rm Name}\\
%Name Institution
} % end author

\maketitle

\begin{abstract}
We evaluate Tencent's QQ Browser, a popular mobile browser in China with
hundreds of millions of users---including 16 million overseas, with respect to
the threat model of a man-in-the-middle attacker with state actor capabilities.
This is motivated by information in the Snowden revelations suggesting that
another Chinese mobile browser, UC Browser, was being used to track users by
Western nation-state adversaries.

Among the many issues we found in QQ Browser that are presented in this paper,
the use of ``textbook RSA''---that is, RSA implemented as shown in textbooks,
with no padding---is particularly interesting because it affords us the
opportunity to contextualize existing research in breaking textbook RSA\@.  We
also present a novel attack on QQ Browser's use of textbook RSA that is
distinguished from previous research by its simplicity.  We emphasize that
although QQ Browser's cryptography and our attacks on it are very simple, the
impact is serious.  Thus, research into how to break very poor cryptography
(such as textbook RSA) has both pedagogical value and real-world impact.

\end{abstract}

\section{Introduction} \label{sec:intro}
Research into ``real-world'' cryptography often focuses on relatively hard
targets, such as SSL/TLS~\cite{drownattack} or
ASP.NET~\cite{Duong:2011:CWC:2006077.2006783}.  Certain market segments,
however, hide much softer targets behind the veil of security-through-obscurity.
Analyzing the cryptography of these self-rolled, lower quality applications of
cryptography has value mainly in terms of impact and pedagogy.  The impact can
be great because, for a variety of reasons ranging from education to market
forces, major software vendors with hundreds of millions of users often
implement very poor cryptography, as we show in this paper.  We present QQ Browser, a browser developed by Chinese company Tencent, as a case study and
identify research opportunities that are specific to the kinds of exploits that
are useful for this type of market segment.  

The pedagogical value of attacking very poor cryptography comes from simplicity.
For example, one of the contributions of this paper is an adaptive
chosen-ciphertext attack (CCA2) on a real-world RSA implementation that can be
easily understood by, \emph{e.g.}, introductory cybersecurity class students.

Through reverse engineering, we have documented the encryption protocols used by
QQ Browser to protect the trove of sensitive information each client uploads to
QQ Browser's servers (this is summarized in Section~\ref{sec:crapto}).  This sensitive
information includes International Mobile Equipment Identifier (IMEI) numbers, web pages
visited, locational data, and many other kinds of private data about a QQ
Browser user.  The possibility that Tencent shares this information with state actors
is explored in existing reports~\cite{jeffqq}, and QQ Browser's data collection mirrors
competing browsers such as UC Browser~\cite{clucbrowser} and Baidu
Browser~\cite{jeffbaidu}.  In this paper we consider a different threat model,
that of attacks by a state actor on QQ Browser's cryptography implementation so that the
state actor does not require Tencent's complicity to violate user privacy.  This is
motivated by reports~\cite{cbcuc,interceptuc} that UC Browser's poor
cryptography was being exploited by the Five Eyes intelligence
agencies\footnote{Australia, Canada, New Zealand, the United Kingdom, and the
United States.} to index that browser's users by IMEI\@.  Note that QQ Browser has
over 16 million users that are outside of China.

One set of vulnerabilities that we present (in Section~\ref{sec:ugly}) would
easily enable indexing of QQ Browser's users by IMEI and decryption of private data
transmitted to QQ Browser's servers.  This set of attacks is based on, \emph{e.g.}, QQ's
poor pseudorandom number generation, use of hard-coded symmetric keys, and use
of a 128-bit RSA key in earlier versions.  These attacks are particularly
devastating, since they would allow any man-in-the-middle attacker, with minimal
resources, to easily decrypt all sessions completely passively and offline.

The second set of vulnerabilities (presented in Section~\ref{sec:bad}) is
centered around QQ Browser's use of textbook RSA\@.  This affords us the opportunity to
contextualize existing research on breaking textbook RSA and present a novel
attack on QQ Browser that is exceptionally simple.  This CCA2 attack allows an attacker
to decrypt any session by making 128 of their own connections to QQ Browser's servers to
crack the session key.  This is a very serious flaw, but would not scale to
indexing all users by IMEI and is not a passive, offline attack.

The third set of vulnerabilities (presented in Section~\ref{sec:notgood}) is
even more severe, because they would, in some cases, allow a man-in-the-middle
attacker to take complete control over a user's device.  We analyze QQ Browser's update
mechanisms for both Android and Windows.  

Our specific contributions are:
\begin{itemize}
\item We demonstrate an extremely simple attack on QQ Browser's pseudorandom number
generator (PRNG) that would enable a state actor, or any other man-in-the-middle
attacker, to easily decrypt any sessions that they were able to record from the
network.  This would be the easiest way to decrypt all sessions offline and
index them by IMEI\@.  We also discuss how previous versions of QQ Browser used
hard-coded symmetric session keys and 128-bit RSA keys.
\item We present an exceptionally simple CCA2 attack on QQ Browser's implementation of
RSA, which is an example of ``textbook RSA'' being used to protect the private
data of hundreds of millions of users.  This attack has pedagogical value
because of its real-world impact and simplicity, and is novel.
\item We re-evaluate Boneh \emph{et al.}'s~\cite{Boneh2000} meet-in-the-middle style attack on
textbook RSA for 128-bit key sizes and modern understandings of state actor
capabilities.  We find that this attack, which would be attractive because it is
passive and offline, does not scale to 128-bit symmetric session keys. 
\item We present man-in-the-middle attacks on the update mechanisms of both the
Android and Windows versions of QQ Browser.  Taken in the context of similar
attacks from previous work, we find that patterns emerge where man-in-the-middle
attackers can develop powerful attack primitives. 
\end{itemize}

Finally, we put these vulnerabilities and related exploits in the context of
related work in Section~\ref{sec:relatedwork}, and find that more research is
needed in certain areas of inquiry to address the problem of poor security and
privacy practices in specific (but very large and important) market segments.
This is followed by a brief summary in the conclusion.

\section{QQ Browser Cryptography} \label{sec:crapto}
When users run QQ Browser on Android, it makes a series of what QQ Browser interally terms as
``WUP requests'' to QQ Browser's server.  These WUP requests contain information such as
a user's International Mobile Equipment Identifier (IMEI), International Mobile
Subscriber Identification (IMSI), QQ username, WiFi MAC address, SSID of
connected WiFi access point and of all in-range access points, Android ID, URLs
of all webpages visited, and other
private information.  More details about WUP requests are available in the
report by Knockel \emph{et al.}~\cite{jeffqq} which analyzes version 6.3.0.1920 of QQ Browser.  Here we focus on the encryption
protocol of the version of QQ Browser that Tencent released as a response to the
vulnerabilities identified in that report.  The main vulnerability that they
fixed that is relevant to the attacks that we present is that they increased the
size of the RSA key from 128 bits to 1024 bits.  Before this fix using
factorization to crack the private key took less than a second on Wolfram Alpha.

Specifically, we analyzed version 6.5.0.2170 of QQ Browser for Android.  This
version, and the updated QQ Browser server, implement the following steps to encrypt WUP
requests from the client to the server:

\begin{enumerate}
\item First, the client generates a 128-bit AES session key for the session,
using a pseudorandom number generator (PRNG) seeded with the current time in milliseconds since the Unix epoch.
\item Then, the client encrypts this session key using a 1024-bit RSA public
key.  The public key has exponent 65537, and the RSA implementation is
``textbook RSA,'' meaning that no form of padding---such as
OAEP~\cite{oaep}---is applied at all.
\item The client uses the AES session key to encrypt the WUP request, in ECB
mode.
\item The client sends the RSA-encrypted AES session key and the encrypted WUP
request to the server.
\item The server decrypts the RSA-encrypted AES key it received from the client
using its private key, then chooses the least significant 128 bits of the
plaintext to be the AES session key.
\item The server decrypts the WUP request using the AES session key that it
obtained \emph{via} RSA decryption.
\item If the AES ciphertext received from the client decrypts to a valid WUP
request correctly, the server sends an AES-encrypted response using the AES
session key (also using ECB mode).
\end{enumerate}

We reiterate the following important points about this protocol because they
will be relevant to the attacks:

\begin{itemize}
\item The only entropy source used by the client to choose the AES session key
is the current time in milliseconds.
\item The client encrypts with the session key first, and the server only
responds if the client's request is properly encrypted with the correct AES key
that the client sent to the server with RSA encryption. 
\item The server ``chops off'' all but the 128 least significant bits of the
decrypted RSA plaintext, with these 128 least significant bits becoming the
128-bit AES session key and all the other bits being ignored.
\end{itemize}

\section{Passive, Offline attacks}
\label{sec:ugly}
In this section, for completeness, we present vulnerabilities in QQ Browser that
are devastating in the sense that they allow all sessions to be decrypted
completely offline.

\subsection{PRNG attack in QQ Browser 6.3.0.1920 for Android} \label{sec:badprng}

\begin{figure*}[th]
\centering
\begin{varwidth}{\linewidth}
\begin{verbatim}
int i = 10000000 + new Random().nextInt(89999999);
int j = 10000000 + new Random().nextInt(89999999);
return (String.valueOf(i) + String.valueOf(j)).getBytes();
\end{verbatim}
\end{varwidth}
\caption{Decompiled Java method generating an AES session key in version
6.3.0.1920.\label{fig:bad}}
\end{figure*}

As described in a previous report~\cite{jeffqq} analyzing QQ Browser version 6.3.0.1920, that version's PRNG algorithm,
shown in in Figure~\ref{fig:bad}, decreases the
entropy of the AES session key used by the client for WUP requests from the
normal $2^{128}$ to $89999999^{2} < 2^{53}$.  This vulnerability was rendered
moot by the fact that this session key is protected by a 128-bit RSA key.

\subsection{128-bit RSA key in QQ Browser 6.3.0.1920}

Although QQ Browser distinguishes itself from competing browsers by attempting to
implement asymmetric cryptography, the RSA key in version 6.3.0.1920 was 128
bits long and is easily factored using Wolfram Alpha or a factorization
library:
\begin{align*}
&245406417573740884710047745869965023463\\
=\:&14119218591450688427 \cdot 17381019776996486069.
\end{align*}

To resolve this vulnerability, QQ Browser increased the RSA key size to 1024
bits in version 6.5.0.2170, but the implementation of RSA is still textbook RSA
with no padding of any kind, leading to attacks on RSA's malleability as
we discuss in Section~\ref{sec:bad}.  Note that even 1024-bit RSA keys are
considered to be too short for modern attackers, and RSA keys of 2048 bits or
more are generally recommended.

\subsection{Hard-coded symmetric keys in QQ Browser 6.3.0.1920}\label{sec:ugly-symmetric}

As described in a previous report~\cite{jeffqq}, QQ Browser version 6.3.0.1920
does use hard-coded symmetric keys in some places, such as
\begin{center}
\texttt{$\backslash$x25$\backslash$x92$\backslash$x3c$\backslash$x7f$\backslash$x2a$\backslash$xe5$\backslash$xef$\backslash$x92}
\end{center}
for DES encryption of the mobile device's WiFi adapter MAC address.

Unlike in version 6.5.0.2170 where the AES session key is used for both WUP
requests and their responses, in version 6.3.0.1920 the session key is only
used for the requests.  Responses from the server are sent using a modified
version of the Tiny Encryption Algorithm (TEA) in a modified CBC block cipher
mode.  The following hard-coded ASCII-encoded key is used:
\begin{center}
\texttt{sDf434ol*123+-KD}
\end{center}

QQ Browser has at least an attempted implementation of asymmetric cryptography, distinguishing it from
less security-conscious browsers such as UC Browser~\cite{clucbrowser} and
Baidu Browser~\cite{jeffbaidu}.

\subsection{PRNG attack in QQ Browser 6.5.0.2170 for Android}

\begin{figure*}[th]
\centering
\begin{varwidth}{\linewidth}
\begin{verbatim}
Random random = new Random(System.currentTimeMillis());
byte[] bArr = new byte[8];
byte[] bArr2 = new byte[8];
random.nextBytes(bArr);
random.nextBytes(bArr2);
return new SecretKeySpec(ByteUtils.mergeByteData(bArr, bArr2), "AES");
\end{verbatim}
\end{varwidth}
\caption{Decompiled Java method generating an AES session key in version
6.5.0.2170.\label{fig:better}}
\end{figure*}

The most recent vulnerability, the attack for which we will refer to as the PRNG
attack, is that the AES key randomly generated for each WUP request is generated
using a random number generator (\emph{java.util.Random}) seeded with the
current time in milliseconds (\emph{java.lang.System.currentTimeMillis()})
before generating every key (this is shown in Figure~\ref{fig:better}.)  Thus,
to guess the key, one must know only the time that the key was generated, which
can be approximated by the time that the WUP request was observed being
transmitted.  By observing a WUP request and using the time of its observation
and simultaneously searching forward and backward from that time, we were able
to guess the correct AES key in fewer than 70,000 guesses, \emph{i.e.}, the key
had been generated within 35,000 milliseconds.

To resolve this vulnerability, QQ Browser can use the
\emph{java.security.SecureRandom} random number generator with no explicitly
passed seed (\emph{i.e.}, with the no-args constructor).

\section{Active Attacks on QQ Browser's Use of Textbook RSA}
\label{sec:bad}
In this section, we explore attacks on QQ Browser's use of textbook RSA.

\subsection{CCA2 attack} \label{sec:cca2attack}

The first attack, which we refer to as the CCA2 attack, results from the fact
that no key padding such as OAEP is used when encrypting the AES key with RSA\@.
Because of this, we are able to leverage the malleability of RSA to perform a
chosen ciphertext attack to guess the AES key one bit at a time.  The threat
model for this attack is an attacker with the ability to record a user's
encrypted session from the network.  We call the client that the user is using
the victim client.  After recording the user's session, the attacker wants to
determine the AES key used for the WUP session so that they can decrypt it.
The attacker accomplishes this by making a series of connections, using its own
client, to the QQ Browser server and attempting encrypted communications with
the server with a series of transformed RSA ciphertexts to gain information
about the original key used by the victim client.

Let $C$ be the RSA encryption of 128-bit AES key $k$ with RSA public key $(n,
e)$.  Thus, we have
\begin{equation*}
C \equiv k^e \pmod{n}
\end{equation*}
Now let $C_b$ be the RSA encryption of the AES key
\begin{equation*}
k_b     = 2^bk
\end{equation*}
\emph{i.e.}, $k$ bitshifted to the left by $b$ bits.  Thus, we have
\begin{equation*}
C_b    \equiv {k_b}^e \pmod{n}
\end{equation*}
We can compute $C_b$ from only $C$ and the public key, as
\begin{equation*}
\begin{split}
C_b &\equiv C(2^{be} \bmod n) \pmod{n} \\
    &\equiv (k^e \bmod n)(2^{be} \bmod n) \pmod{n} \\
    &\equiv k^e 2^{be} \pmod{n} \\
    &\equiv (2^bk)^e \pmod{n} \\
    &\equiv {k_b}^e \pmod{n}
\end{split}
\end{equation*}
The third line follows from the fundamental property of multiplication in modular arithmetic.

We begin the attack by considering $C_{127}$.  It is the RSA encryption of
$k_{127}$, the AES key where every bit but the highest bit are necessarily zero
and where $k_{127}$'s highest bit is $k$'s lowest bit (recall that the QQ Browser server
ignores all but the lowest 128 bits of the decrypted key).  We first guess that
$k_{127}$'s high bit is zero and send a WUP request with $C_{127}$ and encrypt
the request with the key where that bit is zero.  If the server responds, that
means that the bit was zero, since it was able to decrypt our request.  If not,
the bit must have been a one.  After we know this bit, we consider $C_{126}$ and
guess the next bit (note that we know one of $C_{126}$'s bits from $C_{127}$).
We repeat this process for each bit of the AES key.  In total, this requires 128
guesses, since the AES key is 128 bits and each request reveals one bit of the
key.  By using this approach, we can iteratively learn every bit of the AES key.

Recall from Section~\ref{sec:crapto} that the server only responds if the client
sends a properly encrypted WUP request.  If the server sent predictable
plaintext encrypted with the session key from the client without first checking
the client's request to make sure it decrypts properly, we could infer more than
one bit at a time by chopping off, \emph{e.g.}, 16 or 32 bits and performing a
brute-force attack on the plaintext/ciphertext pairs obtained from the server.
However, the client must properly encrypt the WUP request for the server to
respond, so inferring the session key one bit at a time is the most efficient
method of attack, which requires 128 sessions to be initiated with the server by
the attacker.

As discussed in Section~\ref{sec:forgiveus}, we have implemented this attack and
tested it, and informed Tencent of the issue as per ethical disclosure
standards.  To resolve this issue, QQ Browser can use the OAEP key padding algorithm to
encrypt all AES keys.  However, we recommend that they use a well-tested
implementation of SSL/TLS to communicate all WUP requests as this would not only fix
this and other issues (such as the PRNG attack), but also any other undiscovered
issues in their cryptographic implementation.

\subsection{Offline attacks on textbook RSA} \label{sec:factor}

\begin{table}[t!]
    \begin{center}
    \begin{tabular}{|c|c|c|c|}
    \hline
    Bit-length $m$ & $m_1$ & $m_2$ & Probability \\ \hline
    \multirow{4}{*}{64} & 32 & 32 & 17\% \\ \cline{2-4}
    & 33 & 33 & 29\% \\ \cline{2-4} 
    & 34 & 34 & 33\% \\ \cline{2-4}
    & 30 & 36 & 40\% \\ \hline
    \multirow{4}{*}{128} & 64 & 64 & 15\% \\ \cline{2-4}
    & 66 & 66 & 28\% \\ \cline{2-4} 
    & 68 & 68 & 34\% \\ \cline{2-4}
    & 60 & 72 & 39\% \\ \hline
    \end{tabular}
    \caption{Experimental probabilities of splitting into two factors.\label{tab:data}}
\end{center}
\end{table}

The CCA2 attack on QQ Browser is powerful in the sense that a
man-in-the-middle attacker can record a user's session and then easily recover
the session key by testing bits \emph{via} 128 connections to QQ Browser's server.  For
a state actor that wants to decrypt all sessions and index them by IMEI,
however, this is not ideal since over 99\% of the traffic to QQ Browser's server would
be generated by the attacker.  For this reason, we investigated known offline
attacks on textbook RSA, finding that they would not be practical for attacking
QQ Browser.

Boneh \emph{et al.}~\cite{Boneh2000} demonstrate a meet-in-the-middle style
attack on textbook RSA, that is based on the observation that for an encrypted
RSA message $c \equiv M^{e} \pmod{N}$, if we can find small enough integers $M_1 \le 2^{m_1}$
and $M_2 \le 2^{m_2}$ such that $M=M_1 \cdot M_2$, then:

\begin{equation*}
\frac{c}{{M_2}^e} \equiv {{M_1}^e} \pmod{N}
\end{equation*}

By building a table with $2^{m_1+1}\cdot\max(m_1,m_2)$ bits of memory and performing
$2^{m_2}$ modular exponentiations, messages (\emph{i.e.}, session keys) can be
recovered if they can be written as $M=M_1 \cdot M_2$.  The table and search are
per modulus and exponent, so for a single RSA scheme (such as QQ Browser's) the work
would only need to be done once and all sessions could be decrypted.

Table~\ref{tab:data} shows the probabilities that a random 64- or 128-bit
number, $m$, can be written as $M=M_1 \cdot M_2$ such that $M_1 \le 2^{m_1}$ and
$M_2 \le 2^{m_2}$.  The data for 64-bit numbers nearly matches the corresponding
probabilities from Table 1 of Boneh \emph{et al.}~\cite{Boneh2000}, and are only
presented here for verification and comparison purposes.  We generated data for
128-bit numbers because that is the size of QQ Browser's AES session keys.  In terms of
the underlying assumption of Boneh \emph{et al.}'s attack about factoring $M$,
the attack is applicable to QQ Browser's 128-bit session keys.  The resources necessary
to carry out the attack are probably out of the reach of even a state actor,
however.  For example, for $m_1 = m_2 = 64$ and $m=128$, the attack would
require a table of size 295,148 petabytes and $2^{64}$ modular exponentiations.

We discuss Boneh \emph{et al.}'s attack here because we anticipate that in
certain market segments textbook RSA with smaller session key sizes may be
common.  We note that implementations of ElGamal may be susceptible to attacks
for 128-bit session keys, since the attack on ElGamal presented by Boneh
\emph{et al.} can be split into more than two factors.  Also, it may be possible
to combine attacks that reduce the entropy  of the session key with this attack.
Lastly, it may be possible in the CCA2 attack to use RSA's malleability in
combination with Boneh \emph{et al.}'s meet-in-the-middle style attack to hide
from QQ Browser's server which session key is being cracked.

\section{Attacks on QQ Browser's Update Mechanisms for Arbitrary Code
Execution} \label{sec:notgood}

In this section we discuss attacks on QQ Browser's update mechanisms, which are even
more serious than attacks on QQ Browser's cryptography in the sense that any
man-in-the-middle attacker (including state actors) could execute arbitrary code
on a targeted user's machine.  We first discuss a straightforward attack on
the mobile version of QQ Browser's update process.  Then, although we have
discussed only the mobile version of QQ Browser in this paper thus far, we
then discuss the update mechanism in the Windows version of QQ Browser for desktop PCs.
Market share data is unavailable for the Windows version, suggesting that it
has far short of the hundreds of millions of users that its mobile counterpart
has.  However, together with the vulnerability in the mobile version, we
present three increasingly sophisticated attacks on QQ Browser's update process
that demonstrate how digital signature verification of downloaded software is
insufficient to secure an update process against active man-in-the-middle
attacks.

\subsection{Attack on mobile version updates}

The mobile version of QQ Browser checks for and installs updates as follows:
\begin{enumerate}
\item The browser makes a WUP request to the update server containing the
current version of the browser and asking if there are any updates available.
\item The server's response contains a URL to an APK\footnote{An APK is an
Android Application Package, a file format used by the Android operating system
for distributing mobile apps.} and an MD5 hash of the APK
file.  (If no update is available, the server returns a response containing no
update information and the update process halts.)
\item The browser downloads the APK.
\item The browser computes its MD5 hash and verifies it against the one
provided by the server.  (If the hashes mismatch, the browser displays an error
message and the update process halts.)
\item The browser executes the \texttt{ACTION\_VIEW} Android intent against the
downloaded APK.
\end{enumerate}

At this point, the Android operating system takes over.  Under normal
conditions, the system will present a UI asking the user whether to
upgrade QQ Browser to a newer version.  However, other prompts are possible
depending on the APK the browser downloads that a man-in-the-middle attacker
may exploit.  Android requires that an APK upgrading an app be signed with
the same key as that of the currently installed APK, so an attacker cannot
simply upgrade QQ Browser to arbitrary code.  Moreover, Android also does not
allow installing any APK that would downgrade an app, and so a downgrade attack
is not possible.  However, if the downloaded APK is for a different app
than that of QQ Browser or any other app currently installed, then
the user will be prompted to install the APK instead of upgrading QQ Browser.
Although this requires user interaction, most users would be unlikely to
notice or appreciate the significance of being prompted to install a
new package instead of upgrading an existing one, especially if the new package
were designed by an attacker to have the same title and icon of QQ Browser.

In order for a man-in-the-middle attacker to cause the browser to prompt to
install a malicious APK, the attacker must cause the browser to download the
malicious APK and send the browser the corresponding hash.
As the URLs to APKs we observed being sent by the QQ Browser server were all
unencrypted HTTP, a man-in-the-middle attacker could attack the APK download
itself, but then the APK would not have the same MD5 hash as that sent by
the server.  The feasibility of an attacker forging the MD5 hash depends on the
version of QQ Browser requesting updates and the encryption it uses for WUP
requests.

Version 6.3.0.1920 of the browser always receives responses from the server
encrypted with a symmetric, hard-coded key (see
Section~\ref{sec:ugly-symmetric}).  In this version the attacker can respond to
any WUP request to the update server with a forged response containing a
malicious APK URL and its corresponding MD5 hash.

In later versions that use the AES session key to decrypt server responses,
the attack requires a full man-in-the-middle position or a man-on-the-side
attacker who can crack the session key fast enough using the attack in
Section~\ref{sec:crapto} before the browser receives the real server's response.  Alternatively, a
man-on-the-side attacker can have already redirected all traffic via (\emph{e.g.}) DNS
redirection and then perform a man-in-the-middle attack.

\subsection{Attack on Windows version updates}

The Windows version of QQ Browser checks for and installs updates as follows:
\begin{enumerate}
\item The browser sends an unencrypted JSON request to the update server
containing the current version of the browser and asking if there are any
updates available.
\item The server's unencrypted JSON response contains a URL to an
EXE\footnote{Specifically, an EXE is a Windows Portable Executable
(PE) format binary program that can be executed on machines running the Windows
operating system.}, an MD5 hash of an EXE file, and the name of the file to save
the file as.  (If no update is available, the server returns a response
containing no update information and the update process halts.)
\item The browser downloads the EXE and saves it in a temporary directory using
the file name provided by the server.
\item The browser computes its MD5 hash and verifies it against the one
provided by the server.  (If the hashes mismatch, the browser displays an error
message and the update process halts.)
\item The browser verifies the EXE's Authenticode digital signature to ensure
that it was signed by Tencent.  (If it is not, the browser displays an error
message and the update process halts.)
\item The browser executes the downloaded EXE.
\end{enumerate}

\subsubsection{Attack via directory traversal}

Since the update metadata is not protected by any asymmetric cryptography,
a man-in-the-middle attacker can modify any of it.  One attack is possible by
modifying the field specifying the name of the file.  We found that this field
is not sanitized by the browser to prevent directory traversal.  An attacker can overwrite
any file on the user's machine that the user has permission to overwrite.
(Since the file is downloaded before it is verified, it need not have the
correct digital signature nor even be an EXE file.)  For instance, we found
that by using the file name
\urltt{../../../../../../../../../programfiles/tencent/qqbrowser/qqbrowser.exe},
we were able to overwrite the QQ Browser executable with an arbitrary
program.\footnote{Although backslashes are typically used as a path separator on Windows, the Windows kernel generally accepts forward slashes as a path separator as well.}

\subsubsection{Attack via other signed binaries}

We found another vulnerability in the update process that results from the fact
that digital signature verification of an EXE file does not, in general, verify that the
downloaded EXE will perform its intended task such as upgrading the browser.  It only guarantees that the
EXE was signed by Tencent, and so any EXE signed by Tencent can be substituted to satisfy the check.
We found an older web installer for QQ Browser signed by Tencent that downloads
an EXE unencrypted without any digital signature verification.  By first
attacking QQ Browser to download the web installer, and then attacking the web
installer to download a malicious EXE, a man-in-the-middle attacker can still
attack the browser's update process to run an arbitrary program even though the
browser verifies the downloaded program's digital signature.  This attack
requires user interaction to run the web installer, but it is unlikely that a
user would be surprised to have to run an installer after checking for updates.
Moreover, there may exist an undiscovered Tencent-signed executable that
would download and execute code without any required user interaction that
would remove the requirement for user interaction from this attack.

\section{Discussion and Related Work} \label{sec:relatedwork}

Here, we discuss opportunities for research and ethical issues.

\subsection{Opportunities for research}

Although market segments such as Chinese mobile web browsers have very
sophomoric cryptography implementations that lead to very simple attacks, there
are several interesting potential avenues of research.  As pointed out by Bratus
\emph{et al.}~\cite{bratusetal}, an exploit serves as a constructive proof that
``unforeseen computations are indeed possible.''  Exploits also lend credibility
to security concerns and therefore have pedagogical value in relaying the
importance of current best practices (cryptographic or otherwise) to software
developers, policy makers, the public, and others.  Thus, research into
exploiting vulnerabilities in less-developed (in terms of security and privacy)
market segments can have great value.  Here, we point out potential avenues of
research in this respect that are, in our opinion, under-served.

First, we found that there are relatively few attacks in the literature for
textbook RSA\@.  Boneh's survey paper~\cite{Boneh99twentyyears} about attacks on
RSA mostly covers different padding schemes and issues with, \emph{e.g.}, the
choice of public exponent.  Existing CCA2 attacks on RSA
implementations~\cite{drownattack,Duong:2011:CWC:2006077.2006783,woot10,xml,191956}
are all Bleichenbacher-style
attacks~\cite{Bleichenbacher:1998:CCA:646763.706320}.  Two exceptions are Boneh
\emph{et al.}~\cite{Boneh2000} (discussed in Section~\ref{sec:factor}) and
K{\"u}hn~\cite{Kuhn2002}.  The latter presents attacks that are similar to our
CCA2 attack, but for schemes that are de-facto padding schemes.  To the best of
our knowledge, our CCA2 attack is the simplest and possibly the only published
attack for a real implementation of RSA that has no padding.

Second, research into combining PRNG vulnerabilities with other attacks, such as
Boneh \emph{et al.}~\cite{Boneh2000}, could be very valuable for demonstrating
the exploitability of more subtle PRNG issues such as those reported by
Michaelis \emph{et al.}~\cite{Michaelis:2013:RFS:2450083.2450097}.  We
anticipate that the evolution from the current state of cryptography in markets
such as Chinese mobile browsers to current best practices will be a gradual
evolution, and attacks that exploit conversion issues such as
DecryptoCat~\cite{decryptocat} or what we showed in Section~\ref{sec:badprng}
will be very valuable during this transition in order to keep ``raising the
bar.''

Lastly, attacks on update mechanisms could use a more formal treatment to survey
the different attack primitives that are possible.  Buffer overflows and other
memory corruption vulnerabilities have seen considerable research to categorize
different primitives to enable advanced exploit techniques (see, \emph{e.g.},
Bratus \emph{et al.}~\cite{bratusetal} or Shacham~\cite{rop}).  Attacks on
update mechanisms by man-in-the-middle attackers are not new, but they are
becoming increasingly important as state actors build up their capabilities to detect vulnerable update services~\cite{cbcuc,interceptuc} and exploit them
(see, \emph{e.g.},~\cite{191996}).  Patterns emerge when QQ Browser's
vulnerabilities are taken in the context of existing
work~\cite{focimitm,Bellissimo:2006:SSU:1268476.1268483}, such as the re-use of
code signed by a company for other purposes as an exploit primitive.  We believe
that more research in this important area is needed.

\subsection{Ethical considerations} \label{sec:forgiveus}

With the exception of the respective vulnerabilities exploited in our PRNG
attack and CCA2 attack, all vulnerabilities presented in this paper have been
previously published~\cite{jeffqq}, and before that they were subjected to a
45-day vulnerability disclosure process in line with international standards on
vulnerability disclosure~\cite{fortyfivedays}.  We reported the two
vulnerabilities that are newly presented in this paper to Tencent (the
developers of QQ Browser) on 20 April 2016, so this paper is no longer emargoed
as of 4 June 2016.

We tested the CCA2 attack in Section~\ref{sec:cca2attack} against QQ Browser's servers
to verify that it worked.  We cracked session keys for three of our own test
sessions.  By sending QQ Browser's servers ciphertexts that decrypted into plaintexts that went
beyond the 128-bit boundary of a typical session key we were putting QQ Browser's server
at no more than usual risk of denial-of-service than any other public-facing web
server.

\section{Conclusion}
In summary, we have presented three classes of attacks against QQ Browser, a
piece of software that has hundreds of millions of users and collects and
transmits a wide array of private data about them.  The first class of attacks
allowed offline, passive decryption of all sessions recorded.  The second class
of attacks was based on QQ Browser's use of plaintext RSA and included a CCA2 attack
that allowed decryption of targeted sessions \emph{via} 128 active connections
to QQ Browser's servers.  The third class of attacks enabled arbitrary code execution by
a man-in-the-middle attacker.  All three classes of attacks are very serious
and illustrate the importance of further research into attack techniques and
primitives that are common to the emerging threats posed by state actors,
especially in market segments where security and privacy best practices are
underdeveloped.

\section*{Acknowledgments}

This material is based upon work supported by the National Science Foundation
under Grant Nos. \#1518523 and \#1518878.  Kenny Paterson provided useful
insights on attacking QQ Browser's RSA implementation.

\bibliographystyle{acm}
\bibliography{references}

%\theendnotes
%\end{CJK}
\end{document}